\newcommand{\fig}[1]{\mbox{Figure\hspace{0.2em}\ref{#1}}}
\newcommand{\eqn}[1]{\mbox{Equation\hspace{0.2em}\ref{#1}}}
\newcommand{\sect}[1]{\mbox{\S\ref{#1}}}
\newcommand{\tbl}[1]{\mbox{Table\hspace{0.3em}\ref{#1}}}
\newcommand{\dau}{\Delta_\mathrm{AU}}
\newcommand{\NP}{N_\mathrm{P}}
\newcommand{\NT}{N_\mathrm{T}}
\newcommand{\Hs}{H_{l}(D,\Delta)}
\newcommand{\Hsp}{H_{l}}
\newcommand{\vrel}{v_\mathrm{rel}(\phi,\Delta)}
\newcommand{\vrell}{v_\mathrm{rel}(\phi_l,\Delta)}
\newcommand{\vrelp}{v_\mathrm{rel}}
\newcommand{\Nexps}{N_{\mathrm{exp}_l}(D,\Delta)}
\newcommand{\Nexp}{N_{\mathrm{exp}}(D,\Delta)}

\newcommand{\Nexpp}{N_{\mathrm{exp}}}
\newcommand{\nd}{n(D,\Delta)}
\newcommand{\nD}{n_\Delta(D)}
\newcommand{\Es}{E_{l}}
\newcommand{\Rs}{R_{l}(D, \Delta)}
\newcommand{\eff}{\varepsilon_{l}(D, \Delta)}
\newcommand{\Nrec}{N_{\mathrm{rec}_{l}}(D, \Delta)}
\newcommand{\Nadd}{N_{\mathrm{add}_{l}}(D, \Delta)}
\newcommand{\Oe}{\Omega_\mathrm{e}(D, \Delta)}
\newcommand{\Oep}{\Omega_\mathrm{e}}
\newcommand{\ns}{n_\mathrm{s}}
\newcommand{\Ns}{N_\mathrm{s}}
\newcommand{\Ds}{D_\mathrm{s}}
\newcommand{\Dmin}{D_\mathrm{min}}
\newcommand{\Dmax}{D_\mathrm{max}}
\newcommand{\DDD}{dD \, d\Delta}

\documentclass[preprint]{aastex}
\usepackage{graphicx}
\usepackage{natbib}
\usepackage{amssymb}
\usepackage{natbib}

\begin{document}

\title{Upper Limits on the Number of Small Bodies in Sedna-Like Orbits
  by the TAOS Project}

\author{
J.-H.~Wang\altaffilmark{1,2},
M.~J.~Lehner\altaffilmark{1,3,4},
Z.-W.~Zhang\altaffilmark{2},
F.~B.~Bianco\altaffilmark{3,4},
C.~Alcock\altaffilmark{4},
W.-P.~Chen\altaffilmark{2},
T.~Axelrod\altaffilmark{5},
Y.-I.~Byun\altaffilmark{6},
N.~K.~Coehlo\altaffilmark{7},
K.~H.~Cook\altaffilmark{8},
R. Dave\altaffilmark{4},
I.~de~Pater\altaffilmark{9},
R. Porrata\altaffilmark{9},
D.-W.~Kim\altaffilmark{6},
S.-K.~King\altaffilmark{1},
T.~Lee\altaffilmark{1},
H.-C.~Lin\altaffilmark{2},
J.~J.~Lissauer\altaffilmark{10},
S.~L.~Marshall\altaffilmark{8,11},
P. Protopapas\altaffilmark{4},
J.~A.~Rice\altaffilmark{7},
M.~E.~Schwamb\altaffilmark{12},
S.-Y.~Wang\altaffilmark{1} and
C.-Y.~Wen\altaffilmark{1}
}
\altaffiltext{1}{Institute of Astronomy and Astrophysics, Academia Sinica.
  P.O. Box 23-141, Taipei 106, Taiwan}
\email{jhwang@asiaa.sinica.edu.tw}
\altaffiltext{2}{Institute of Astronomy, National Central University, No. 300,
  Jhongda Rd, Jhongli City, Taoyuan County 320, Taiwan}
\altaffiltext{3}{Department of Physics and Astronomy, University of
  Pennsylvania, 209 South 33rd Street, Philadelphia, PA 19104}
\altaffiltext{4}{Harvard-Smithsonian Center for Astrophysics, 60 Garden Street,
  Cambridge, MA 02138}
\altaffiltext{5}{Steward Observatory, 933 North Cherry Avenue, Room N204
  Tucson AZ 85721}
\altaffiltext{6}{Department of Astronomy, Yonsei University, 134 Shinchon,
  Seoul 120-749, Korea}
\altaffiltext{7}{Department of Statistics, University of California Berkeley,
  367 Evans Hall, Berkeley, CA 94720}
\altaffiltext{8}{Institute for Geophysics and Planetary Physics, Lawrence
  Livermore National Laboratory, Livermore, CA 94550}
\altaffiltext{9}{Department of Astronomy, University of California Berkeley,
  601 Campbell Hall, Berkeley CA 94720}
\altaffiltext{10}{Space Science and Astrobiology Division 245-3,
  NASA Ames Research Center, Moffett Field, CA, 94035}
\altaffiltext{11}{Kavli Institute for Particle Astrophysics and Cosmology,
  2575 Sand Hill Road, MS 29, Menlo Park, CA 94025}
\altaffiltext{12}{Division of Geological and Planetary Sciences, California
  Institute of Technology, 1201 E. California Blvd., Pasadena, CA 91125}

\begin{abstract}
We present the results of a search for occultation events by objects
at distances between 100 and 1000~AU in lightcurves from the
Taiwanese-American Occultation Survey (TAOS). We searched for
consecutive, shallow flux reductions in the stellar lightcurves
obtained by our survey between 7 February 2005 and 31 December 2006
with a total of $\sim4.5\times10^{9}$ three-telescope simultaneous
photometric measurements. No events were detected, allowing us to set
upper limits on the number density as a function of size and distance
of objects in Sedna-like orbits, using simple models.
\end{abstract}

\keywords{Kuiper Belt -- occultation -- Solar System: formation}

%%%%%%%%%%
\section{Introduction}

During and just after the epoch of planet formation, a significant
number of planetesimals were left that had not accreted into planets.
Some of the planetesimals near the giant planets were ejected to orbits
with large semi-major axes, and many more were hurled hyperbolically
into interstellar space. Those objects which formed beyond
proto-Neptune would probably remain near the ecliptic plane and became
part of the Kuiper Belt \citep{1950BAN....11...91O,
  1974CeMec...9..321K, 1980Icar...42..406F, 1987AJ.....94.1330D}.

Up to the present, more than 1000~Kuiper Belt Objects (KBOs) have been
discovered\footnote{See
  \url{http://www.cfa.harvard.edu/iau/lists/TNOs.html} for a list of
  these objects.}. The estimated total mass of the Kuiper Belt from
observations to date is $\sim 0.1~$M$_{\oplus}$, which is a few
orders of magnitude less than the mass predicted for the minimum mass
solar nebula \citep{1977Ap&SS..51..153W}.  Theoretical simulations
show that in order to assemble 1000-km size objects between 30 and
50~AU, initially there must have been much more mass in this region
\citep{2004Natur.432..598K, 2005AJ....129..526S} than the current
observations suggest. Many surveys indicate a Kuiper Belt ``cliff''
beyond 50~AU \citep{1999AJ....118.1411C, 2001AAS...199.6310A,
  2004AJ....128.1364B}. Given that more than 99\% of the original disk
mass appears to have been depleted, some sort of dynamical
perturbations must have removed objects from this region.

Furthermore, the origin of $2003~$VB$_{12}$ (Sedna, with orbital
parameters of $a$=531~AU, $q$=76~AU, $i=12^\circ$) is puzzling
\citep{2004ApJ...617..645B}. Searches for thermal radiation from Sedna
with the IRAM 30~m telescope and Spitzer have put an upper limit on
its diameter of $1600$ km \citep{2004DPS....36.0301B}. The mass of
Sedna, from the upper limit on the size and mean density of KBOs, is
estimated to be $\leq 10^{-3}~$M$_{\oplus}$. Given its large size and
highly eccentric orbit, in situ formation seems unlikely.  Sedna is
unique because its perihelion at 76~AU is far outside the reach of
dynamical perturbation by Neptune, and the aphelion of $\sim$1000~AU
is also too small for Sedna to be affected sufficiently by giant
molecular clouds or galactic tides in the solar
neighborhood. Meanwhile, recent observations
\citep{2005A&A...439L...1B,2007A&A...466..395E} have shown that Sedna
is likely to have methane, water and nitrogen ices.  The similarity of the
spectroscopic features between Sedna and Triton
\citep{2005A&A...439L...1B} suggests its formation near the giant
planet region.

Several theories have been proposed to explain the existence of Sedna
in its present orbit. For instance, a close stellar encounter
\citep{2000Icar..145..580F, 2000ApJ...528..351I, 2004Natur.432..598K,
  2004AJ....128.2564M} could have depleted the Kuiper Belt and excited
outer Solar System objects, such as Sedna, to their present
orbits. Other possibilities have also been suggested, including rogue
mars-sized bodies, or multiple stellar encounters in the Sun's birth
cluster \citep{2004ApJ...617..645B, 2004AJ....128.2564M,
  2006ApJ...643L.135G, 2007Icar..191..413B}. 

Note that all of these theories also predict that many more objects
would be found in \emph{Sedna-like} orbits, i.e. orbits with perihelia
large enough to be unaffected by Neptune and aphelia small enough to
be unaffected by galactic tidal forces or giant molecular clouds.
Furthermore, the fact that Sedna was discovered very close to
perihelion, where it spends a very small fraction of its orbital
period, led \citet {2004ApJ...617..645B} to infer that there may be as
much as many as 500 objects in similar orbits with the same size as 
Sedna or larger.
However, recent search attempts \citep{2007AJ....133.1247L,
  2008ssbn.book..335B, schwamb2009} did not find any new member of
this population. In their recent survey, \citet{schwamb2009} modeled
objects with the same semi-major axis and eccentricity as Sedna.  A
best fit of 40 objects brighter than or equal to Sedna in that region
is consistent with their results of detecting one object with a
perihelion past 70~AU.

Most objects in Sedna-like orbits would have escaped from direct
detection due to their enormous distances from the Earth.  Direct
detection of this kind of object is limited to relatively close
objects and by large telescopes, so the majority of the population is
still well beyond the detection limit for most ground-based
telescopes. \citet{1976Natur.259..290B} suggested that the existence
of the ``invisible'' objects in this region might be inferred by
stellar occultation, and the utility of using robotic telescopes to
search for occultations by outer Solar System objects has been
demonstrated \citep{axelrod1992, Cook1995}.

The Taiwanese-American Occultation Survey (TAOS) aims to investigate the 
size distribution of small ($\sim$1~km) KBOs \citep{2009PASP..121..138L,
  2008ApJ...685L.157Z}.  Four 50~cm robotic telescopes have been
installed at Lulin mountain in central Taiwan.  Each telescope is 
a fast F/2 system with a 3 deg$^{2}$ field of view, equipped with a 
2k$\times$2k CCD camera that can perform 5~Hz
photometric sampling on a few hundred stars simultaneously. Multiple
telescopes are used to reduce the false positive event rate.  The TAOS  
system has been in routine operation with three telescopes since
February 2005, and four-telescope observations began in August 2008.

TAOS was designed to search for KBOs near 50~AU, for which a typical
occultation by a KBO of a few km across would manifest itself as a
flux reduction in only one or two consecutive time series
measurements.  With a null detection, a stringent constraint has been
set on the number and size distribution of KBOs by
\citet{2008ApJ...685L.157Z}.  However, the same data set can be used
to detect objects in the distant Solar System all the way to
$\sim$1000~AU; i.e., the survey is also sensitive to objects in
Sedna-like orbits.  For objects at 100 -- 1000~AU, the Fresnel scale
and the projected size of the background star are larger than for
objects in the Kuiper Belt, resulting in occultation events with
longer durations \citep{2007AJ....134.1596N}. Thus a distant
occultation event would have flux reductions spanning \emph{several}
lightcurve points and requires a different analysis pipeline on the
TAOS data from that reported in \citet{2008ApJ...685L.157Z}.  Here we
present our analysis of the TAOS data to search for objects on
Sedna-like orbits.  An estimate of the number of such objects detected
would help constrain the population of bodies beyond 50~AU, thereby
confronting theoretical models of the early Solar System, and address
the puzzle of mass deficit in the region.

% Figure 1
\begin{figure*}[b]
\plottwo{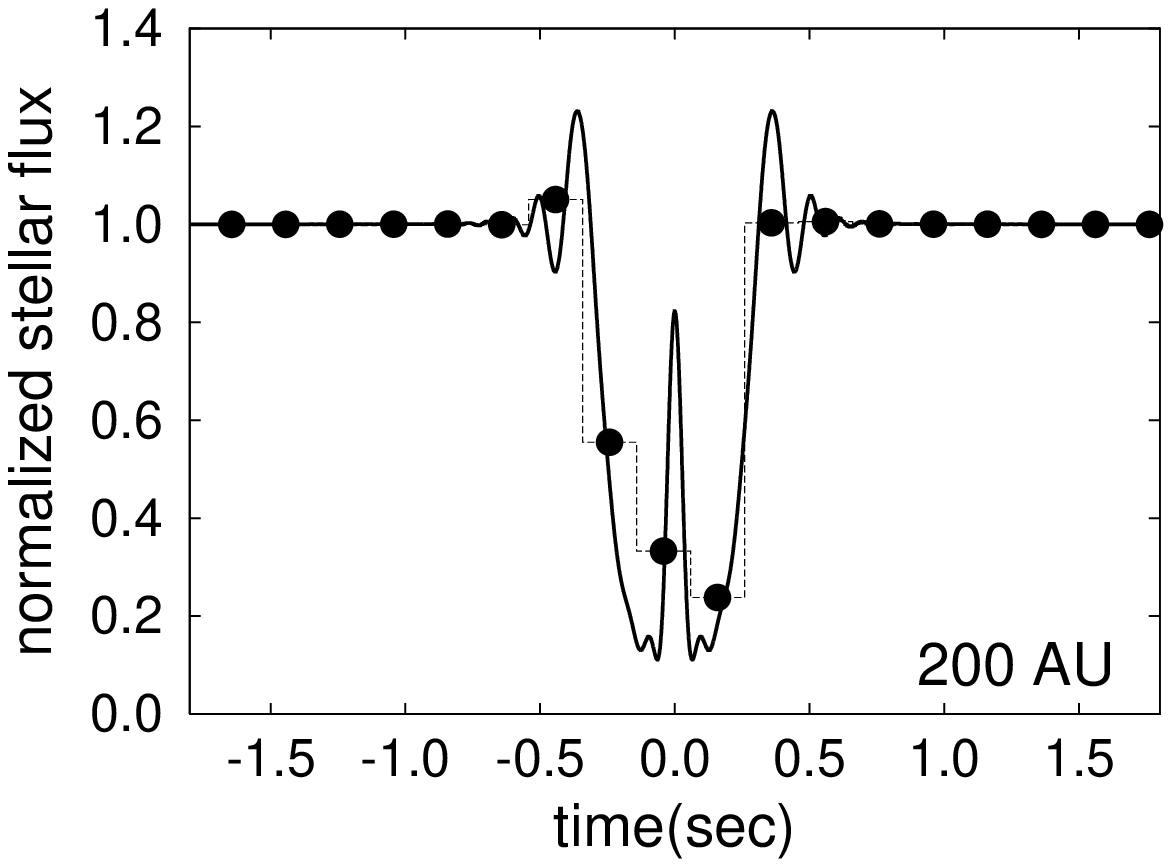}{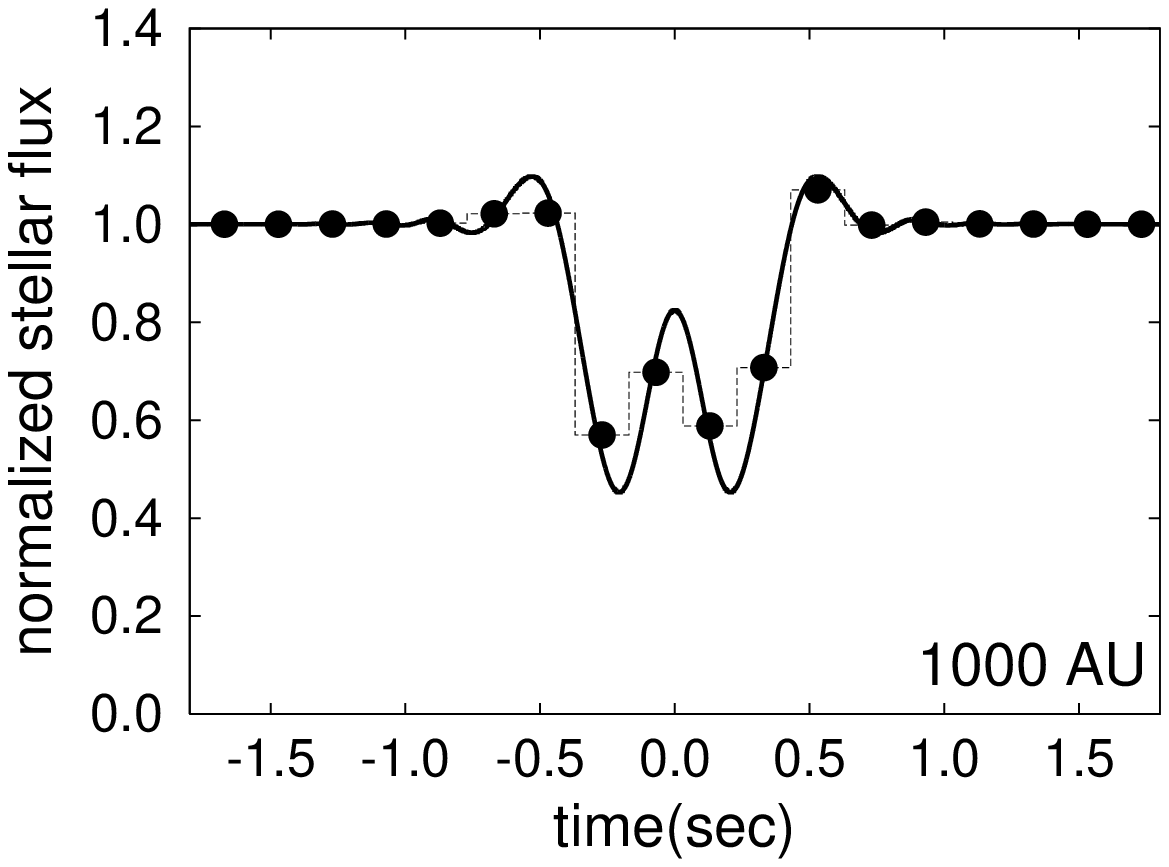}
\caption{Diffraction profiles for 10~km diameter object occulting
  a 0.01 mas background star (projected stellar diameter $\sim7$~km at
  1000~AU) at 200~AU and 1000~AU. Solid lines are the diffraction
  profiles, solid circles represent 0.2~sec (5~Hz) sampling light
  curve points, dotted lines indicate the duration of the 0.2 sec
  integration. The sampling starts at random time offset. The
  velocity of the object was calculated at an
  opposition angle of $30^{\circ}$.}
\label{fig:diffraction}
\end{figure*}

%%%%%%%%%%%%%%
\section{Detection Algorithm}
\label{sec:det}

Stellar occultations by Solar System objects fall into the domain of
Fresnel diffraction \citep{1980poet.book.....B, 1987AJ.....93.1549R}
if the object diameter $D \lesssim F$, the \emph{Fresnel scale}, which
is given by
\begin{displaymath}
F = \sqrt{\lambda \Delta/2},
\end{displaymath}
where $\lambda$ is the observing wavelength and $\Delta$ is the
distance from the object to the observer.  For the TAOS system, where  
$\lambda \sim 600$~nm, this corresponds to objects with $D \lesssim
2$~km at 100~AU or $D \lesssim 7$~km at 1000~AU.  A high cadence
system like TAOS will translate diffraction profile into several 
time-sampled lightcurve points every second as the shadow of a stellar 
occultation sweeps across a telescope.

\fig{fig:diffraction} shows examples of diffraction profile for a
10~km diameter object occulting a 0.01~milli-arcsecond (mas)
background star (V=11 A0V star) at 200 and 1000~AU, 
with a zero impact parameter
(defined as the line-of-sight distance between the centers of object
and background star).  Note that we integrated every 0.2~sec with a
random starting time to simulate the 5~Hz sampling of the TAOS
observations. In order to find possible events in the TAOS
lightcurves, our task is to detect such shallow flux reductions in
consecutive lightcurve points.

\begin{figure}[b]
\plotone{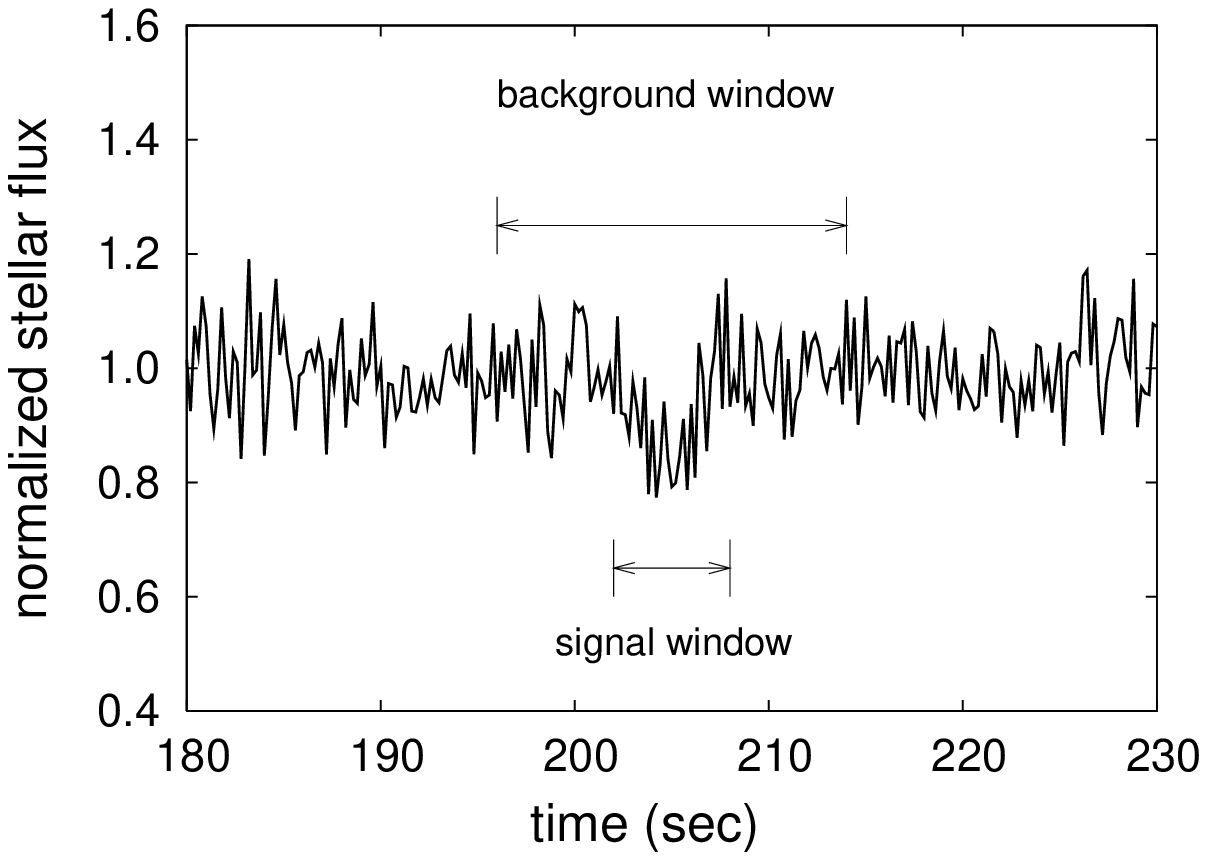}
\caption{In this simulated lightcurve, the local background level is
  estimated in the background window. After subtracting the nominal level,
  the summed value within the signal window, i.e. the EW index, is
  used to search for a possible occultation event. }
\label{fig:ewwindow}
\end {figure}

To detect occultation events by objects in Sedna-like orbits, we have
applied an algorithm to identify consecutive, relatively shallow flux
drops. The key is to recognize systematic, weak flux reduction with an
unknown duration. For each point in the lightcurve, we defined two
windows consisting of a set of consecutive photometric measurements
centered on that point.  First, a window (\emph{background} window) of
a certain size is selected, in which the median value serves as the
local nominal flux level. Then, at the center of the background
window, a smaller window (\emph{signal} window) is chosen that is
relevant to the size and distance of the objects we want to search for
(see \fig{fig:ewwindow}).  The sum of deviations of all data values
from the local nominal value within the signal window then is a
measure of systematic brightening or dimming in the lightcurves.  A
large cumulative deviation occurring simultaneously in multiple
telescopes suggests the possibility of an occultation event within the
signal window.  The concept is similar to the equivalent width (EW) in
spectroscopy for the strength of a spectral line relative to local
continua and is similar to what \citet{2006AJ....132..819R} used in their
work. We define the EW index as
\begin{equation}
W_{i} \equiv
 \sum_{j=i-m}^{i+m}\left(\frac{f_{j}}{\left\langle f\right\rangle } -1\right),
\end{equation}
where $\langle f\rangle$ is the median value from all the measurements
within the background window, and $f_{j}$ is the flux of the
$j^\mathrm{th}$ lightcurve point within the signal window $\{i-m,
i+m\}$ centered at $i$.  The EW algorithm works because flux
reductions of consecutive lightcurve points, albeit weak, accumulate
to result in a small EW index.  To search for events, these windows
were centered at each point in the lightcurve set\footnote{We define a
  lightcurve set as a set of lightcurves of a single star taken
  synchronously with multiple telescopes.}, except the beginning and
end of a lightcurve set where we cannot define a complete background
window, and the corresponding $W_{i}$ value was calculated.
 
The size of the signal window is determined by the occultation
event width and the relative velocity between the observer and the 
object. The relative velocity is estimated as
\begin{equation}
  \vrel = v_\oplus\left(\cos\phi - \sqrt{\frac{1}{\dau}
    \left(1-\frac{1}{\dau^2}\sin^2\phi\right) }\right),
 \label{eq:vrel}
\end{equation}
where $v_\oplus=29.8$~km/sec is the orbital speed of the Earth,
$\dau$ is the distance from the object to the observer in units of AU,
and $\phi$ is the opposition angle defined as angle from opposition to
the line of sight to the background star. \citet{2007AJ....134.1596N}
gave an approximation of the occultation event width as
\begin{equation}
  H=\left[ (2 \sqrt{3} F(\lambda,
    \Delta))^{\frac{3}{2}}+(D)^{\frac{3}{2}}\right]^{\frac{2}{3}}+
       \Delta \, \theta_{*}
\end{equation}
where $D$ is the diameter of the object, and $\theta_{*}$ is the
angular size of the occulted star.  The size of the signal window is
given by $u=sH /\vrelp$, in which $s = 5$~Hz is the TAOS sampling
rate.  As an example, $u=5$ for a 0.05~mas star and 3~km diameter
object at 300~AU and $\phi=30^{\circ}$.  We have confirmed empirically
that the choice of the width of the background window is not critical,
as long as it is sufficiently larger than the signal window but not so
large as to average over any slowly varying trends. A typical value is
about four times the size of the signal window.

% Figure 2
%%%%%%%%%%

A noticeable negative EW index indicates a significant local cumulative flux drop
over the signal window, possibly due to an occultation event. However,
with single telescope data, one cannot rule out the possibility that
this signal has been caused by some random or other unknown
processes. The advantage of a multiple-telescope system, like TAOS, is
the ability to discriminate against such false positive events.

% Figure 3
\begin{figure*}[b]
\plottwo{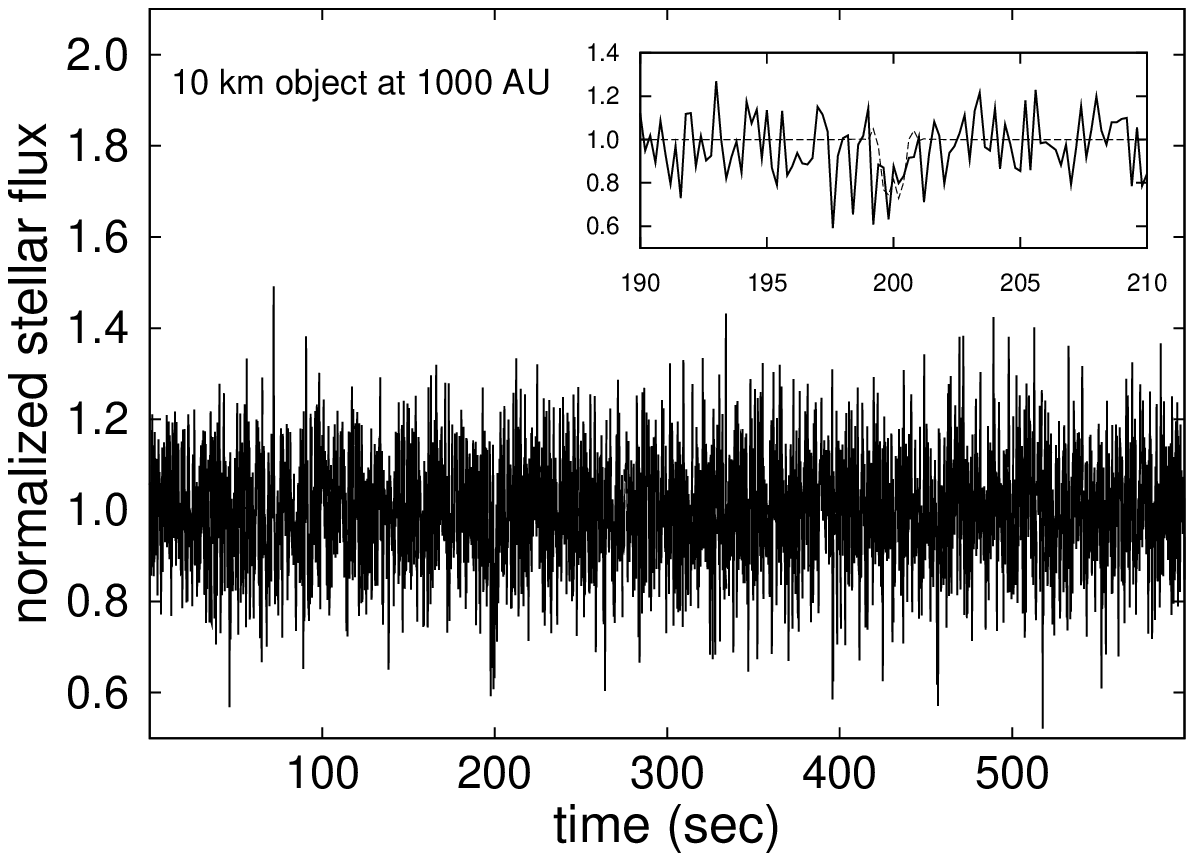}{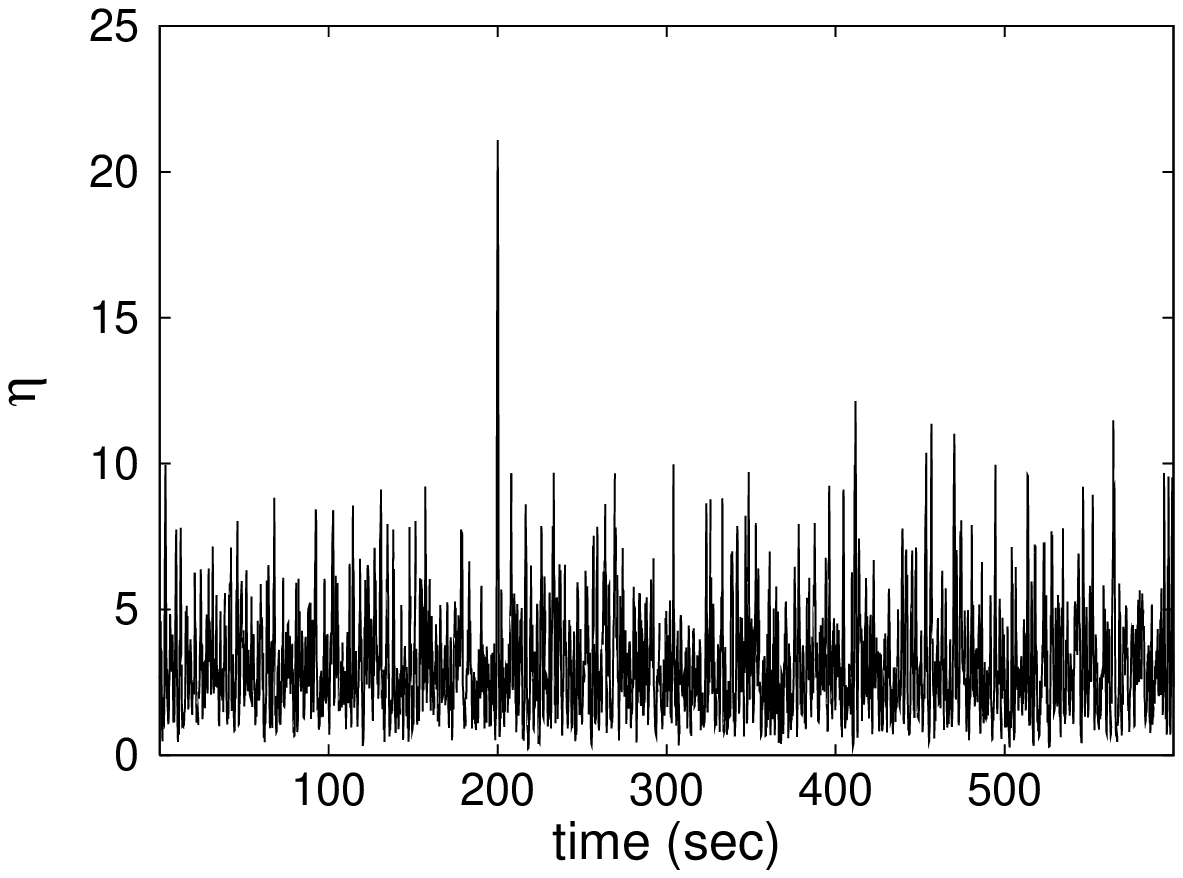}
\caption{The weak flux drop for consecutive lightcurve points (inset,
  left panel) is not readily discernible in the \emph{single} simulated
  lightcurve (we plot only 3000 out of 27000 lightcurve points from
  one telescope for clarity), however, it stands out clearly as a very
  strong signal after processing simulated multi-telescope data with
 the EW~algorithm, as shown in the right panel.}
\label{fig:simewrank}
\end{figure*}

We identify simultaneous low EW indices by \emph{ranking} the EW
indices in each telescope from the lowest to highest, that is, in a
single lightcurve, the lowest EW measurement in the time series would
have a rank $r = 1$, and the largest EW index would have rank $r =
\NP$, where $\NP$ is the total number of EW indices from each
telescope. We then calculate the parameter

\begin{equation}
  {\eta_{i}=-\ln\prod_{k=1}^{\NT}\left(\frac{r_{ik}}{\NP}\right)},
\label{eq:eta}
\end{equation}

where $r_{ik}$ is the rank of $W_{i}$ from telescope $k$, and $\NT$ is
the total number of telescopes.\footnote{This is analogous to the
approach used in \citep{2008ApJ...685L.157Z}.}
If the EW indices are low on each
telescope at a single time point, the rank product should also be very
low, giving rise to a large value of $\eta$. This is illustrated in
\fig{fig:simewrank}. The left panel shows a small section of a long
\emph{simulated} lightcurve (27,000 lightcurve points for a typical
TAOS observing run) with an occultation by a 10~km diameter object at
1000~AU with $\phi = 30^{\circ}$.  An occultation was implanted at $t
= 200$~seconds, and is not obvious at all in any of the single
lightcurves.  However, after processing with the EW algorithm on the
three lightcurves, the event stands out clearly (\fig{fig:simewrank},
right), demonstrating the power of multi-telescope data and our
analysis pipeline. 
% Figure 4
\begin{figure*}[b]
\plotone{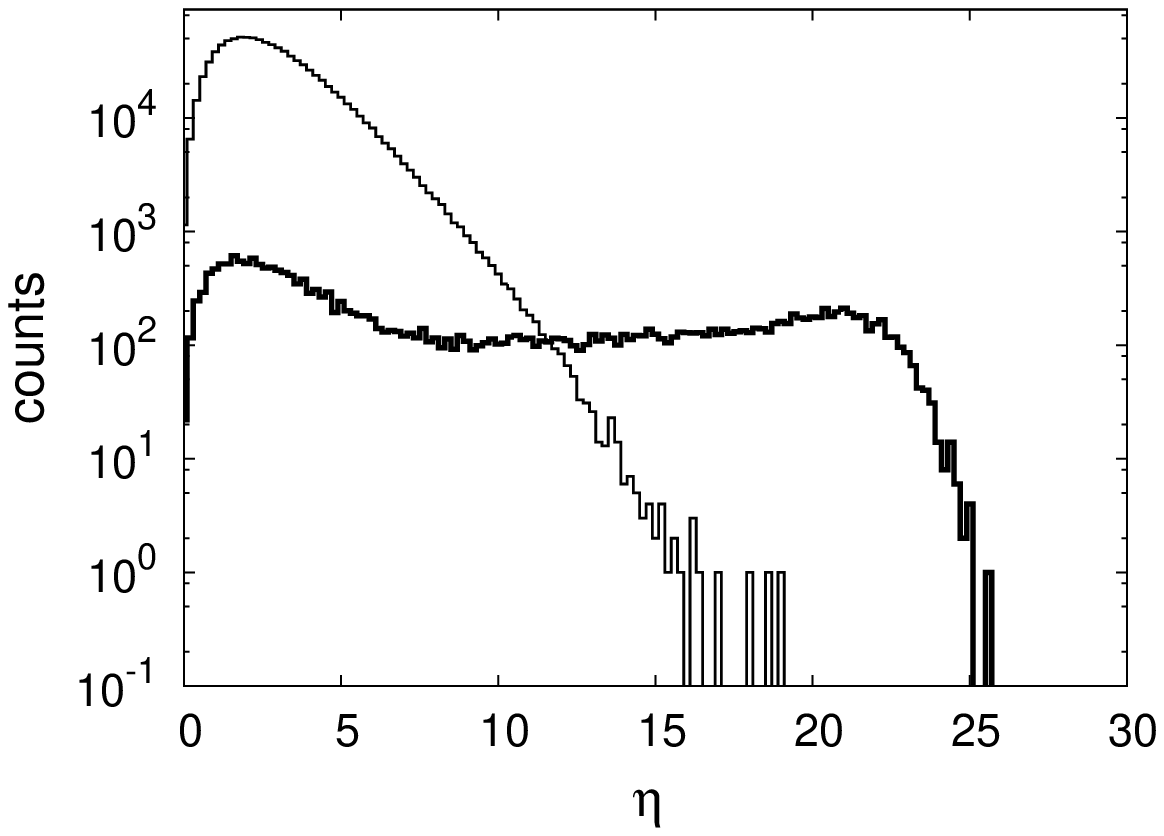}
\caption{Thin line: distribution of $\eta$ from a data run with no
  occultation events. Thick line: values of $\eta$ where one synthetic event
  was added to each lightcurve set and the test was repeated 22139
  times to get the distribution. The synthetic
  events are from 5~km diameter object at 300~AU.}
\label{fig:gammadist}
\end{figure*} 
The values $r_{ik}$ are uniformly distributed and discrete over $\{1,
\NP\}$. If we make the approximation that the values are continuous,
it can be shown \citep[see][for example]{rice} that if there is no correlation
between the data from different telescopes, and if the variance of the
EW index in a filtered lightcurve remains constant throughout the
duration of a data run\footnote{We define a data run as a series of
  three-telescope observations of a particular star field.}, the
distribution of $\eta$ follows the $\Gamma$-distribution of the form
\begin{equation}
  p(\eta)=\frac{\eta^{{\NT-1}}}{\Gamma(\NT)} \, e^{-\eta}.
\end{equation}

This is illustrated in \fig{fig:gammadist}, which shows a histogram of
$\eta$ for a single data run, along with a histogram of $\eta$ where one
synthetic event was added to each lightcurve set and the test was
repeated 22139 times to get the distribution. While many of the synthetic
events are marginal and have low values of $\eta$, there is clearly a
significant population of events at large values of $\eta$. 
To set a cut on the parameter $\eta$, we note that $\eta$ depends
strongly on the value of $\NP$, the length of the lightcurve
set. Since, as discussed in \sect{sec:proc}, this value can vary
significantly due to data runs which are shortened by bad weather, we
can derive a parameter which would be valid for all data runs by
integrating the tail of the distribution to calculate

\begin{eqnarray}
F(\eta) & = & \int\limits_{\eta}^{\infty}
d\eta^\prime\frac{\eta^{\prime{\NT-1}}}{\Gamma(\NT)} \,
e^{-\eta^\prime}\nonumber\\ & = & \frac{\gamma(\NT,
  \eta)}{\Gamma(\NT)}.
\label{eq:gammaint}
\end{eqnarray} 

In this paper we use $F(\eta)$ to select possible events.  $F(\eta)$ is
well-suited to this application, as illustrated in  \fig{fig:gammadist}
which shows the histogram for a data run in which there are no 
evident events, and also a histogram of $F(\eta)$ for the same 
data run, but with artificial events inserted into the same data.

We can choose a threshold value of $F$ that will give a 
few (if any) candidate events yet still ensure a
sufficiently high event detection efficiency. We thus set a threshold
of $F \leq 10^{-8}$, which was chosen empirically after analysis of a
subset of data. The validity of this approach is demonstrated in
\sect{sec:eff}, where we show that our detection algorithm gives a
significant detection efficiency in the absence of any actual detected
events.

With further analysis we will be able, in principle, to use this approach
to put a rigorous upper bound on the expected false positive rate.
We do not do this in this paper, and the absence of detected events,
as shown below, legitimizes our decision.  We determine our detection
efficiency (see Section 4) by direct simulation, and thus translate our
absence of detections into a formal upper limit on the populations
of objects with orbits similar to Sedna's.

%%%%%%%%%%%
\section{Processing The TAOS Data}
\label{sec:proc}
A customized pipeline using aperture photometry was developed to
process the TAOS image data. This process is described in detail in
\citet{photpaper}, but a short summary is given below.

At the beginning of every data run, a series of stare-mode images is
collected. These images are used to find the position of target stars
using \emph{SExtractor} \citep{1996A&AS..117..393B}. Then these stars
are cross-matched with USNO-B1 star catalog
\citep{2003AJ....125..984M} and the position of all stars in the image
with TAOS instrumental magnitude $M_\mathrm{TAOS}<13.5$ are
extracted. An aperture mask is then created for each star identified
in the stare mode image. The aperture size for each target star is
optimized by the signal-to-noise ration (SNR) of the lightcurves from
first 1000 photometric measurements, and the aperture size that gives
the highest SNR is chosen and used.

\begin{figure*}[b]
\plottwo{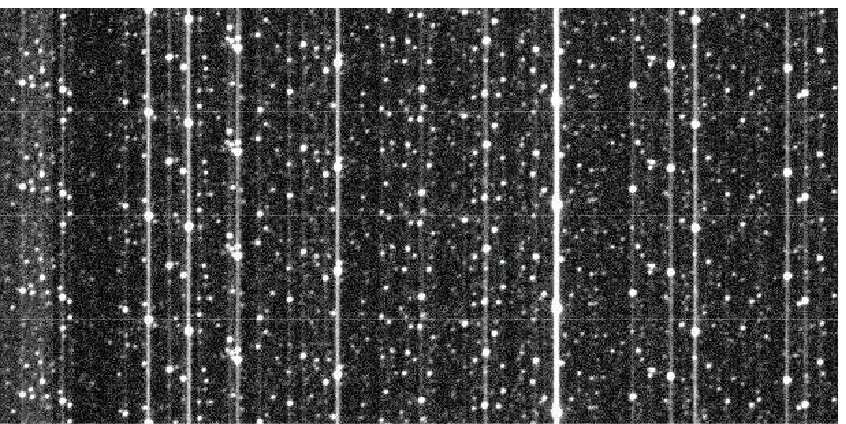}{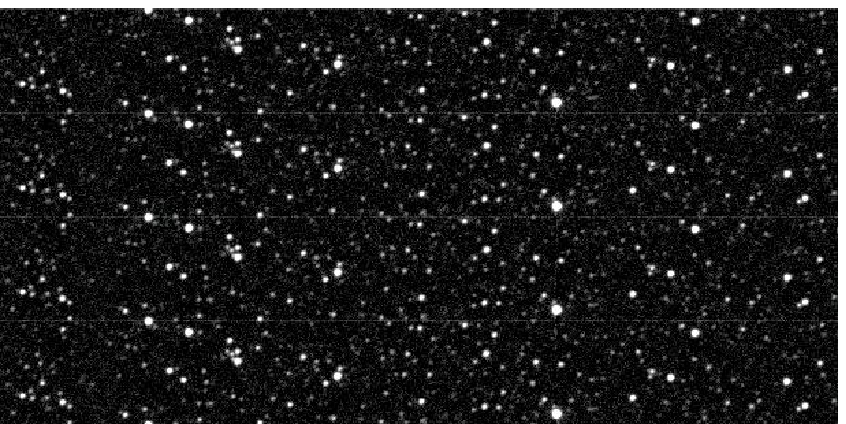}
\caption{The Background and streak removal was done  
by subtracting the median value of each column which effectively 
removed the background and streaks left behind by bright stars.}
\label{fig:streak}
\end{figure*}

The first step in analysis of the time series zipper mode images is
the subtraction of the sky background and the streaks from the
brighter stars \citep[see][for a description of zipper mode imaging
  and the origin of the streaks in the images]{2009PASP..121..138L}.
This was done by subtracting the median value of each column which
effectively removed the background and streaks
simultaneously. \fig{fig:streak} shows a subsection of a raw zipper
mode images and the same image after the background and streak
subtraction.

The next step in the photometry process is to apply the aperture mask
and add the photon counts. The position of the masks was adjusted at
each timestamp to correct for any offset due to tracking errors. The
offset is calculated by calculating the centroids of 20~bright stars
collected at each epoch in the data run.

During the photometry pipeline process, various flags were assigned to
reflect known effects that would cause faulty photometric
measurements, for example, for stars that are near or at the CCD image
boundary or with incorrect exposure times due to system timing
errors. These errors were all recorded and the corresponding
lightcurve points were flagged and removed.

% FIGURE 5

% FIGURE 6
\begin{figure*}[b]
\plotone{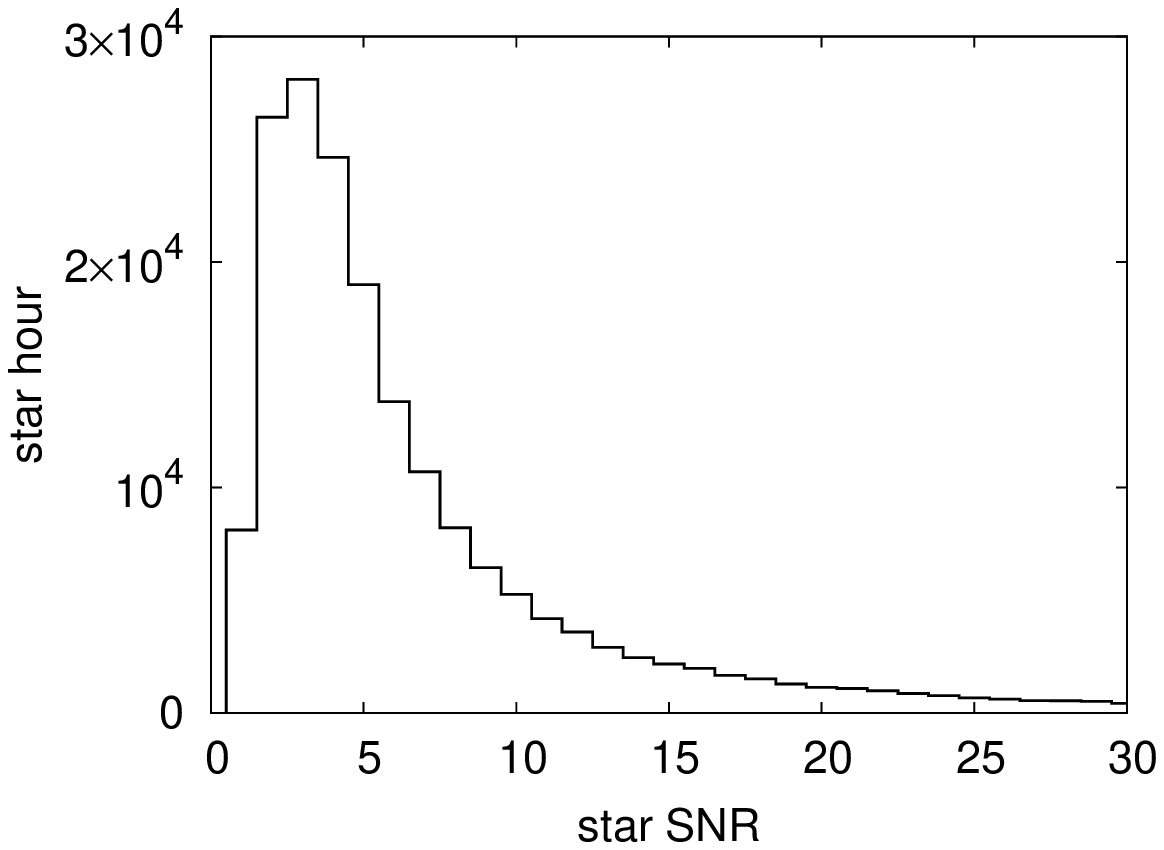}
\caption{The histogram of SNR from all lightcurves with more than
  10,000~data points.}
\label{fig:snr}
\end{figure*}

A complete TAOS data set from 2005 February 7 to 2006 December 31,
which amounts to $\sim 4.5 \times 10^{9}$ three-telescope measurements
was used to search for possible events by objects in Sedna-like
orbits. Only measurements from observing runs with more than 10,000
measurements were included in the analysis, because short duration
runs could not provide long enough baseline to derive significant
statistics. (The duration of a normal data run is 90~minutes, however,
occasional interruptions by system failures or bad weather would
prevent the system from taking data with the normal duration.)
Removing these data runs leaves a total of 199 data runs which amounts
to $1.7 \times 10^{5}$~star-hours. A plot of the number of star-hours
versus SNR for this data set is shown in \fig{fig:snr}.  Furthermore,
an additional 26~data runs were found to have significant correlations
among the lightcurves for each star
\citep{2008ApJ...685L.157Z}. Removing these runs leaves us with
173~data runs, corresponding to a total of $1.7 \times 10^{5}$~star
hours.

Meanwhile, some of the earlier observing runs suffered from guiding
problems in the beginning of each run. They all happened near the
first 20 seconds when the guider tried to adjust the tracking for the
first time. The common feature for this problem is a large
displacement in both RA and Dec coordinates of every star. Hence, the
first minute of data runs with these large positional offsets were
removed.

\begin{figure*}[b]
\plottwo{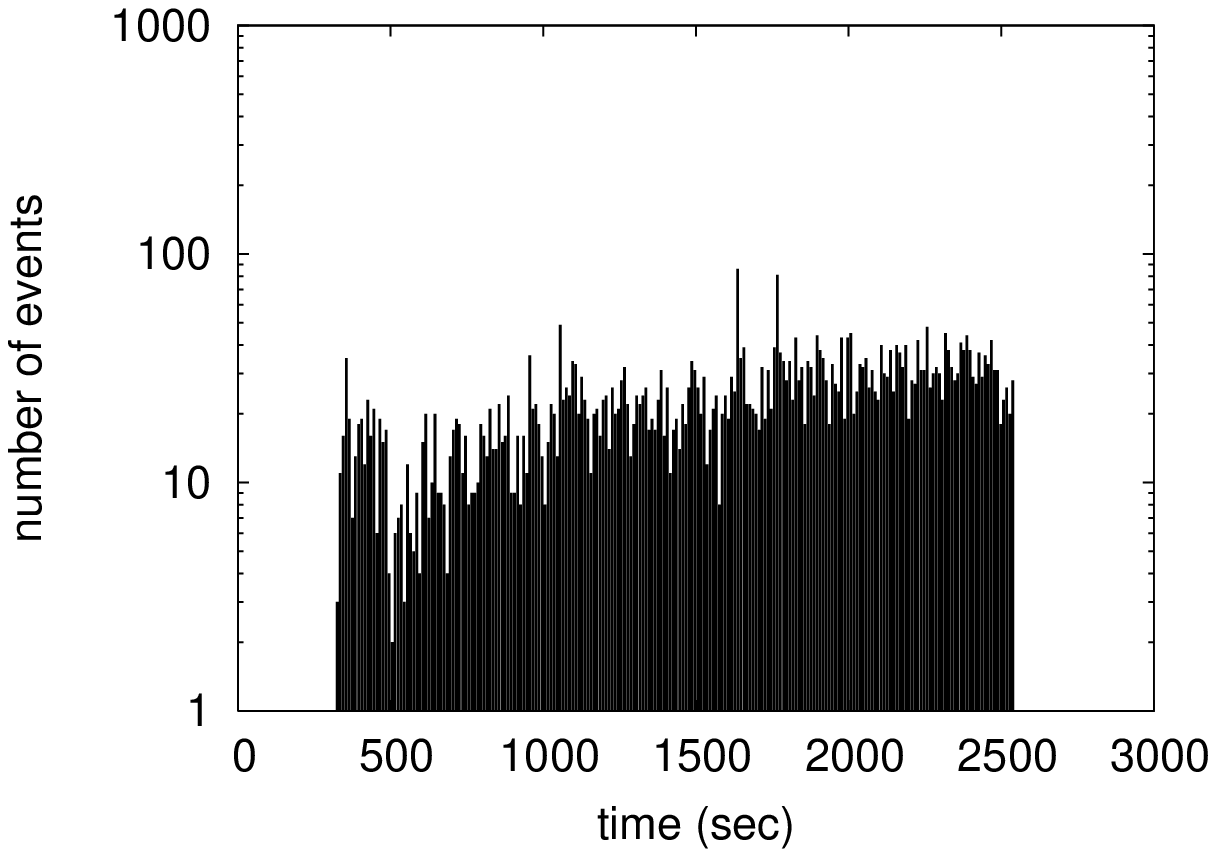}{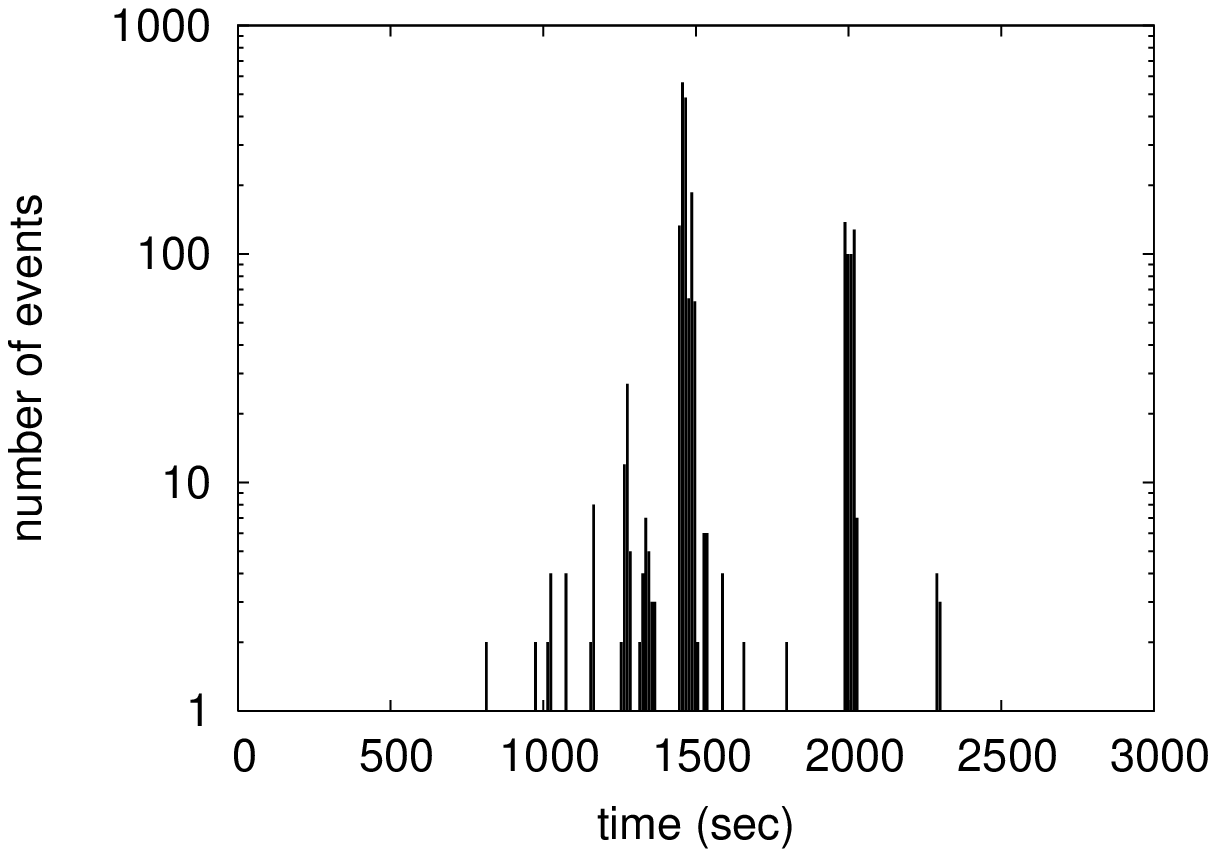}
\caption{Histograms of the number of candidate events vs. time, with
  ten second binning. A run with possible high altitude clouds or
  cirrus (left) would cause frequent flux reduction to many stars
  within a short period of time. Many time coherent events could be
  observed throughout the entire run. On the other hand, in a run with
  moving clouds (right), only sporadic coherent events happened
  at the time when the clouds moved into the our target field. }
\label{fig:histo}
\end{figure*}

After all bad lightcurve points were removed, the EW filter was
applied to the remaining lightcurves. Any rank triplets beyond our
threshold of $F \leq 10^{-8}$ are flagged as candidate events. Most of
the candidate events identified by the EW~filter turned out to be
\emph{time coherent}.  \fig{fig:histo} shows two such cases of the
number of candidate events within a short period of time (ten
seconds).  For example, a run with possible high altitude clouds or
cirrus would cause overwhelming flux reductions for most stars at the
same time. Data runs exhibiting such effects are removed from further
analysis. A total of 31~data runs were removed by this cut. On the
other hand, during a run with low-lying moving clouds, coherent events
would happen only sporadically, when the clouds cover some of the
stars in the target field. Data that show such time dependent
patterns, which are well characterized and understood, would not be
analyzed further for event detection.

As a final selection criterion, we required all candidate events to
have at least three consecutive flux measurements below the nominal
background level $\langle f \rangle$ used to calculate $W_i$. In the
end, no candidate occultation events from objects in Sedna-like orbits
were found in the first two years of the TAOS data.

% Figure 7

\section{Efficiency Calculation and Event Rate}
\label{sec:eff}
For a given lightcurve set $l$, the differential occultation event rate
$R_{l}$ as a function of diameter $D$ and distance $\Delta$ is
\begin{equation}
\frac{d^{2} \Rs}{\DDD}=\frac{d^{2}\nd}{\DDD}\frac{\vrell\Hs}{\Delta^{2}},
\end{equation}
where $\nd$ is the number density of objects with diameter $D$ at
distance $\Delta$, $\vrell$ is the relative velocity between occulting
object and observer during the data run (see \eqn{eq:vrel}), and $\Hs$
is the event width, which is assumed to be the diameter of the first
Airy ring of the diffraction profile as discussed in
\sect{sec:det}. (The subscript on $\phi_l$ indicates that the relative
velocity depends on the angle between the target star and opposition
during the data run, and the subscript on $H_l$ indicates that the
event cross section depends on the angular size of the target star.)

With the event rate, we can now calculate the number of events 
expected to be seen for a \emph{given star} during a given data run, as
\begin{equation}
  \frac{d^{2} \Nexps}{\DDD} = \frac{d^{2}\Rs}{\DDD} \,  \Es \, \eff,
  \label{eq:8}
\end{equation}
where $\Es$ is the duration of the data run, and $\eff$ is the
\emph{efficiency} for which occultations by objects of diameter $D$ at
distance $\Delta$ are detected in the lightcurve set during the
data run.

The efficiency $\eff$ was estimated by a simulation, in which we
implanted occultation events in the actual light curves, and processed
with the same EW filter.  If we added $\Nadd$ events with diameter $D$
and distance $\Delta$ to a lightcurve set and recover $\Nrec$ of them,
the detection efficiency is simply $ \eff = \Nrec/\Nadd$.  Even though
the impact parameter and epoch at which an occultation takes place
affect if a particular event can be detected, by choosing these
parameters randomly over their true distributions, we automatically
average the efficiency over these parameters, and they can be
ignored. We thus chose the impact parameters uniformly in the interval
$[-\Hsp/2, \Hsp/2]$, and the event epoch uniformly in the interval
$[t_0, t_0 + \Es]$, where $t_0$ is the start time of the data run.

With $\eff$ known, the total number of events expected to be seen in the 
entire data set can be estimated from \eqn{eq:8}.  That is, 
\begin{equation}
   \frac{d^{2}\Nexp}{\DDD} = \sum_{l} \frac{d^{2}\Rs}{\DDD} \, \Es \, \eff,
\end{equation}
where the sum is taken over all lightcurve sets in the TAOS data set.

Given the extremely large number of lightcurve sets, we found we could
calculate a statistically significant value of the detection
efficiency by adding \emph{only one} simulated event to each
lightcurve set. If we add events with diameter $D$ to a fraction $w_D$ of all
lightcurve sets, and events at distance $\Delta$ to a fraction $w_\Delta$
of all lightcurve sets, we can write
\begin{displaymath}
  \Nadd = w_D w_\Delta.
\end{displaymath}
We thus choose a set of diameters $D$ and a set of corresponding
values of $w_D$ such that $\sum_D w_D = 1$, and a second set of
distances $\Delta$ and corresponding values of $w_\Delta$ such that
$\sum_\Delta w_\Delta = 1$.

Now there would be only one event in a given lightcurve set, so,
taking into account of the weighting factors of $w_D$ and $w_\Delta$,
we have the total number of expected events, expressed in the fully
expanded form,
\begin{equation}
 \frac{d^{2}\Nexp}{\DDD} = \frac{d^{2}\nd}{\DDD} \frac{1}{w_D
   w_\Delta}\sum_{l}^\mathrm{rec} \frac{\vrell\Hs\Es}{\Delta^{2}},
\end{equation}
where the sum is now over all lightcurve sets where an event of
diameter $D$ and distance $\Delta$ is recovered.  This expression can
be simplified as \citep{2008ApJ...685L.157Z},
 \begin{equation}
 \frac{d^{2} \Nexp}{\DDD} = \frac{d^{2}\nd}{\DDD} \, \Oe,
   \label{eq:nexp}
\end{equation}
where
\begin{equation}
 \Oe = \frac{1}{w_D w_\Delta}\sum_{l}^\mathrm{rec}
 \frac{\vrell\Hs\Es}{\Delta^{2}}
\end{equation}
is the \emph{effective solid angle} of the survey. 

\eqn{eq:nexp} shows that the number of expected events $\Nexpp$ in our
survey depends on two factors, a \emph{model dependent} factor $n$ and
a \emph{survey dependent} factor $\Oep$.  
This indicates that the number of events detected in the
survey and the value of $\Oep$ determined by the efficiency calculation
can thus be used to place constraints on $\nd$.

\begin{figure*}[]
\begin{center}
\begin{tabular}{cc}
\hspace{-1cm}
\resizebox{80mm}{!}{\includegraphics{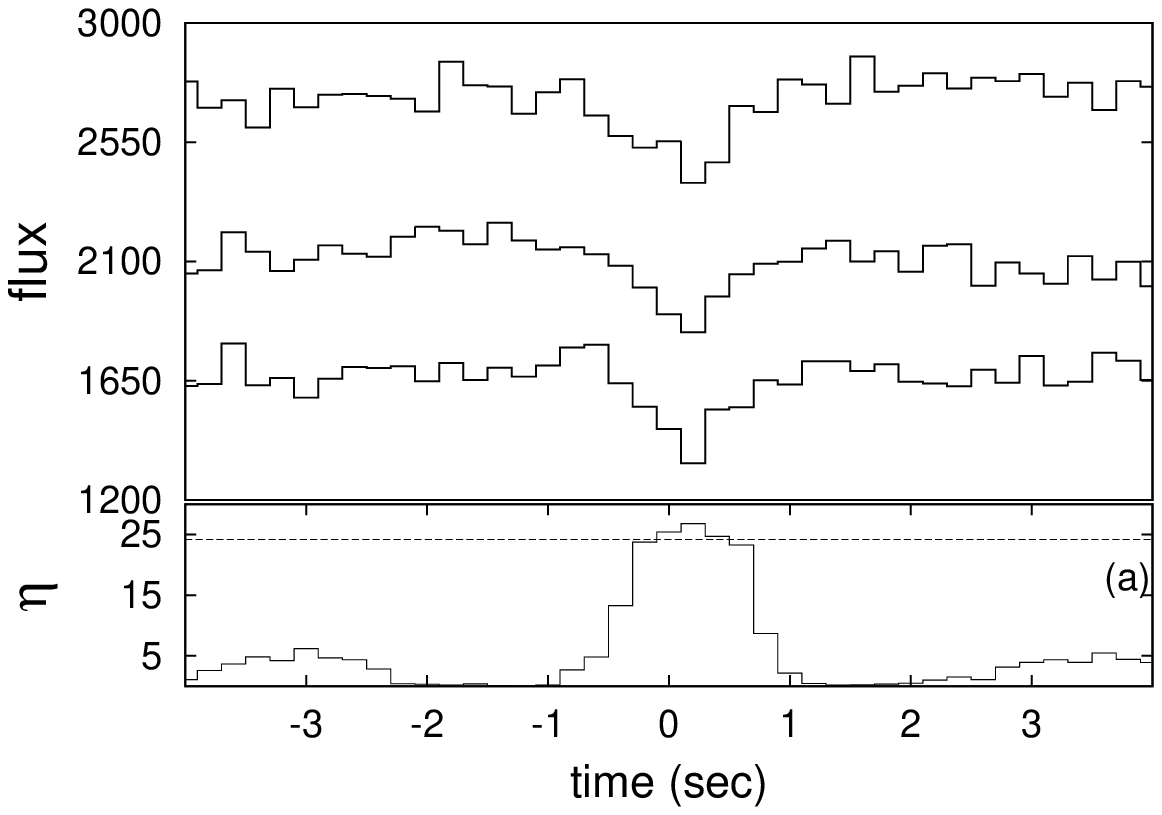}} &
%\hspace{-1cm}
\resizebox{80mm}{!}{\includegraphics{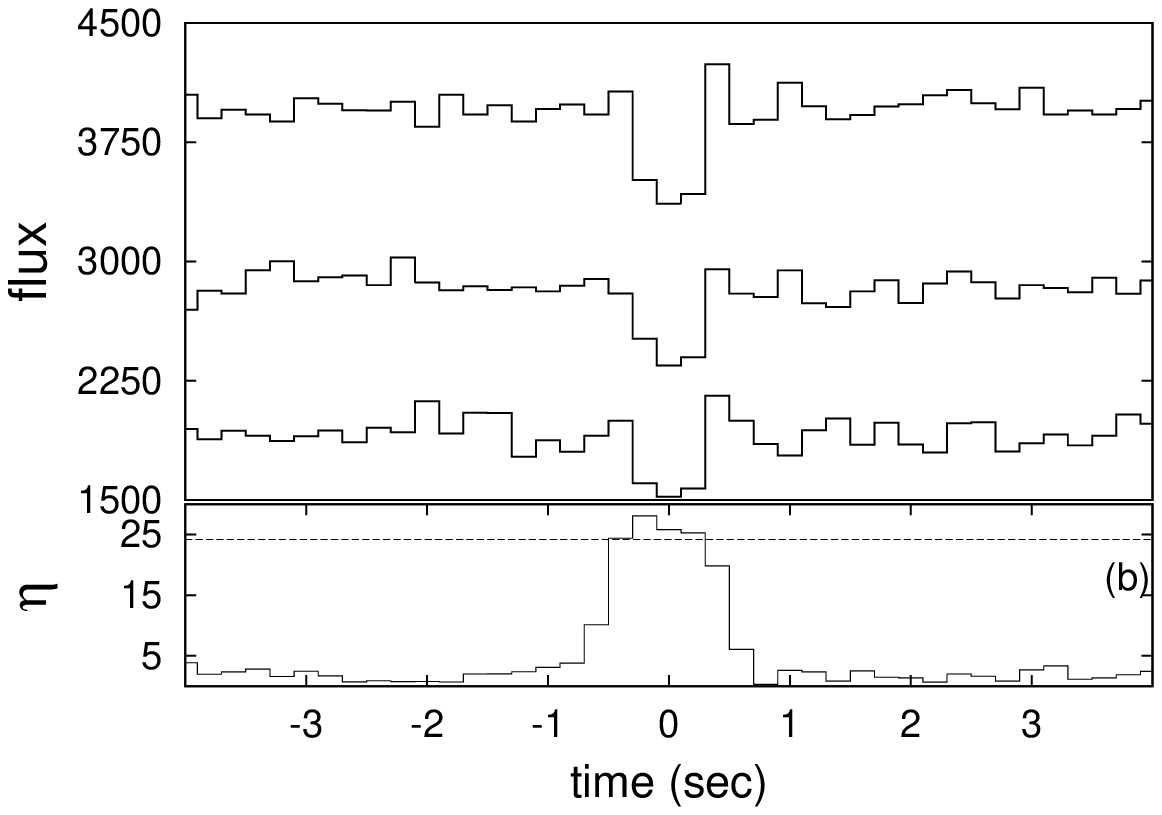}} \\
\hspace{-1cm}
\resizebox{80mm}{!}{\includegraphics{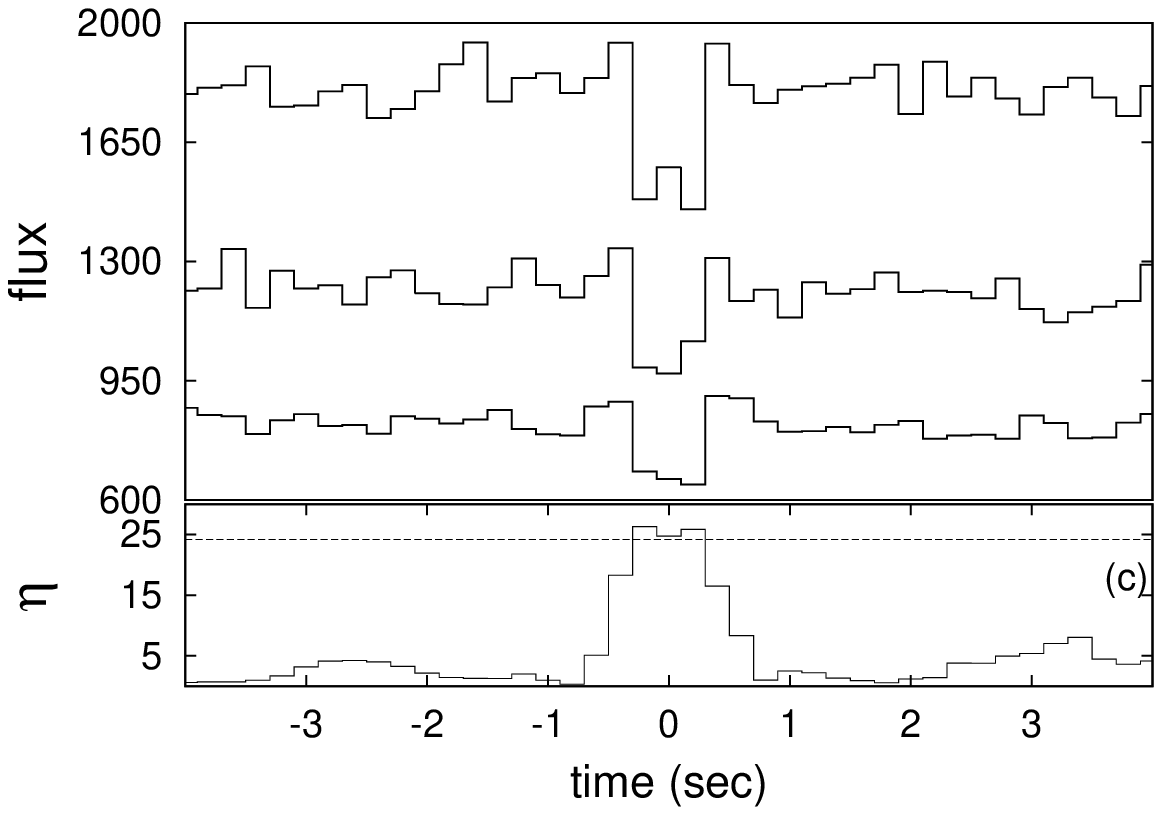}} &
%\hspace{-1cm}
\resizebox{80mm}{!}{\includegraphics{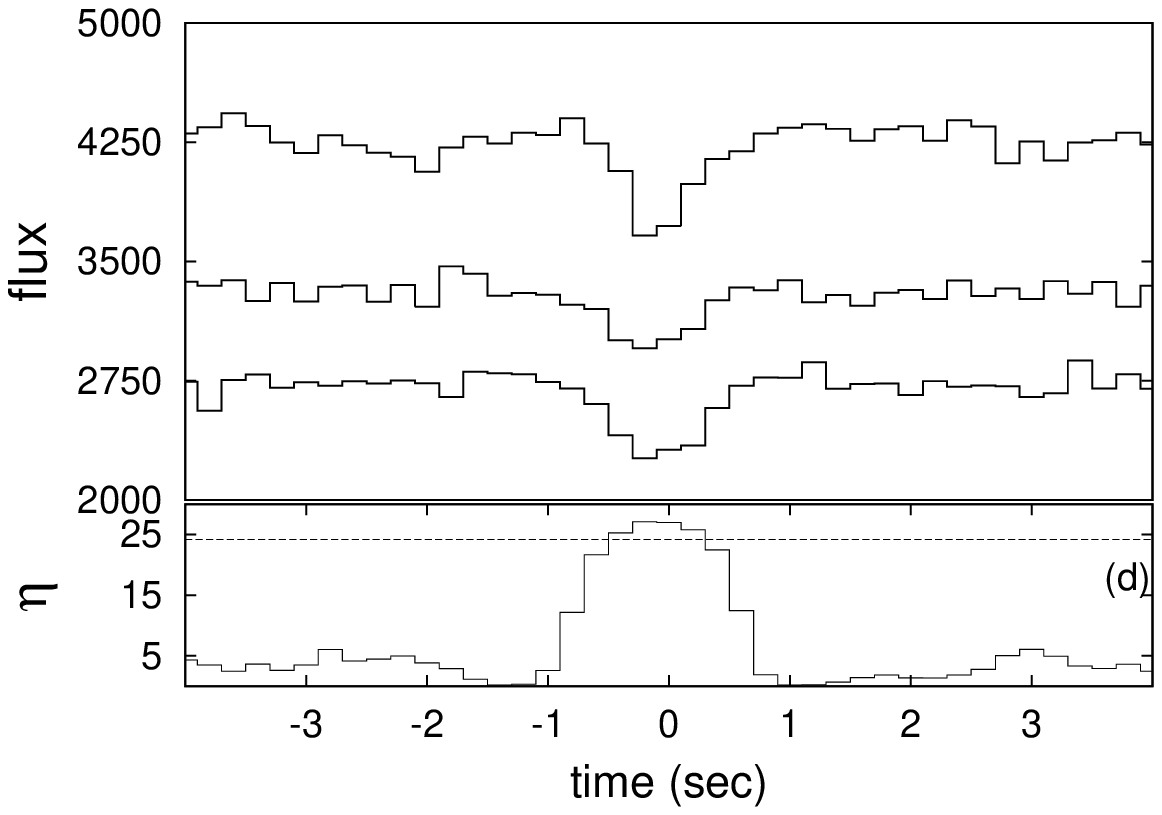}} \\
\end{tabular}
\caption{During the efficiency test,  synthetic events were added into TAOS lightcurves. Examples of lightcurves with  the synthetic events in three TAOS lightcurves are shown in the upper part of each panel and 
the corresponding $\eta$ values  in the lower part of each panel.
 Some constants were added to second and third lightcurves to separate them for clarity. 
 The dotted-lines are the $\eta$ value for $F \le 10^{-8}$.  (a) 2~km, 200~AU, $\phi=68.8^{\circ}$ (b) 3~km, 500~AU, $\phi=45.9^{\circ}$, (c) 5~km, 1000~AU, $\phi=34.4^{\circ}$ (d) 10~km, 1000~AU, $\phi=34.4^{\circ}$.}
\label{fig:eventlc}
\end{center}
\end{figure*}

\begin{figure}[b]
  \plotone{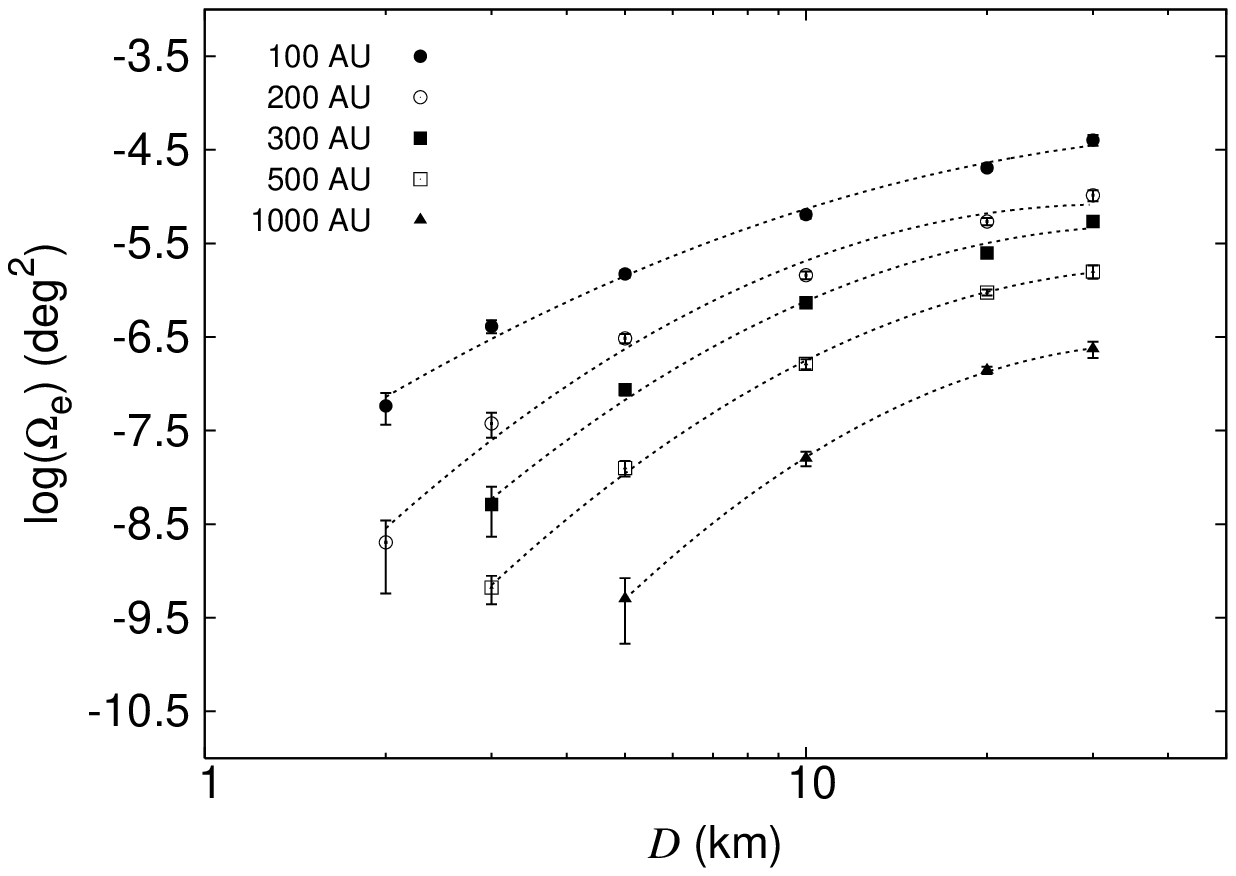} 
  \caption{Effective solid angle versus size for different distances
    from the TAOS efficiency calculation.}
  \label{fig:aeff}
\end{figure}

As in \citet{2008ApJ...685L.157Z}, we used the event simulator
described in \citet{2007AJ....134.1596N} in our efficiency
calculation.  The distances, diameters, and weighting factors used are
shown in \tbl{tab:weights}. The weights were chosen to give preference
to smaller objects where the detection efficiency is expected to be
low in order to improve the statistical accuracy of $\Oep$. After
implanting simulated events, the exact same selection criteria
described above were applied to find which events are
recovered. Several example lightcurve sets where simulated events are
recovered are shown in \fig{fig:eventlc}, and the resulting values of
$\Oep$ are shown in \fig{fig:aeff}.

% Table 1
\begin{deluxetable}{rrrr}
\tabletypesize{\small}
\tablecolumns{4}
\tablewidth{0pc}
\tablecaption{Distances, diameters, and weighting factors used in
  efficiency test. The weighting factors are normalized so that
  $\sum_D w_D = 1$ and $\sum_\Delta w_\Delta = 1$.}

\tablehead{$D$ (km) & $w_D$ & $\Delta$ (AU) & $w_\Delta$}
\startdata
0.5  & 100/785 & 100  & 0.2\\
0.7  & 100/785 & 200  & 0.2\\
1.0  & 100/785 & 300  & 0.2\\
1.3  & 100/785 & 500  & 0.2\\
2.0  & 100/785 & 1000 & 0.2\\
3.0  & 100/785 &  & \\
5.0  & 100/785 &  & \\
10.0 & 50/785  &  & \\
20.0 & 30/785  &  & \\
30.0 & 5/785   &  & \\
\enddata
\label{tab:weights}
\end{deluxetable}

% Table 2
\begin{deluxetable}{rrrrrr}
\tabletypesize{\small}
\tablecolumns{5}
\tablewidth{0pc}
\tablecaption{Distribution of synthetic events in
  efficiency test}
\tablehead{$D$ (km) &100~AU & 200~AU & 300~AU & 500~AU & 1000~AU}
\startdata
0.5 & 0/6409 & 0/6325 & 0/6496 & 0/6361 & 0/6378 \\
0.7 & 0/6393 & 0/6339 & 0/6392 & 0/6420 & 0/6411 \\
1.0 & 0/6339 & 0/6363 & 0/6466 & 0/6365 & 0/6453 \\
1.3 & 0/6260 & 0/6391 & 0/6291 & 0/6357 & 0/6281 \\
2.0 & 21/6403 & 3/6432 & 0/6382 & 0/6285 & 0/6458 \\
3.0 & 68/6408 & 23/6366 & 5/6375 & 1/6427 & 0/6296 \\
5.0 & 207/6302 & 96/6402 & 57/6357 & 21/6398 & 2/6318 \\
10.0 & 291/3116 & 203/3227 & 172/3285 & 86/3246 & 29/3221 \\
20.0 & 327/1897 & 291/1927 & 257/1936 & 193/2010 & 96/1855 \\
30.0 & 78/295 & 66/311 & 69/366 & 51/331 & 26/310 \\
\enddata
\label{tab:implant}
\end{deluxetable}

The ratio of recovered to implanted events for each value of $D$ and
$\Delta$ is shown in \tbl{tab:implant}. Note that the algorithm did
not recover any events with $D < 2$~km. Also note that the efficiency
for detecting 30~km objects at 100~AU is only about 30\%, indicating
that a large fraction of the low SNR lightcurve sets used in the
analysis are not useful for detecting occultation events at these
distances. While including these stars in the analysis lowers the
detection efficiency, they have no effect on the resulting
values of $\Oep$. However, in future analysis runs we will likely set
stricter cuts on SNR values to use our computing resources more
efficiently.

% FIGURE 8

% Figure 9

Given $\Oep$ and the fact that no events were found by the survey, we
can set constraints on $\nd$. Because there are no published models of
the size and distance distributions of objects in this region, we
place constraints on $\nd$ at each value of $\Delta$ listed in
\tbl{tab:weights}.  For the size distribution, we assume a simple
power law model of the form
\begin{equation}
 \frac{d \,\nD}{d\,D}=n_{0}(q-1)\left(\frac{D}{1~\mathrm{km}}\right)^{-q},
 \label{eq:dndD}
\end{equation}
where $q$ is the slope of the power law, and $n_0$ is
the number density of objects with $D \ge 1$~km at distance
$\Delta$. The expected number of events at distance $\Delta$ is thus
\begin{equation}
N_{\rm exp}(\Delta) = n_0\,(q-1)\int^{\Dmax}_{\Dmin}
\left(\frac{D}{1~\mathrm{km}}\right)^{-q}\Oe \, dD,
\end{equation}
where $\Dmin$, for each distance, is the detection limit of size below which  
the efficiency goes to zero.  We picked $\Dmax=30$~km
because we expected the number density of larger objects to be so
small as to provide a negligible contribution to the event rate.

% Figure 10
\begin{figure}[b]
\plotone{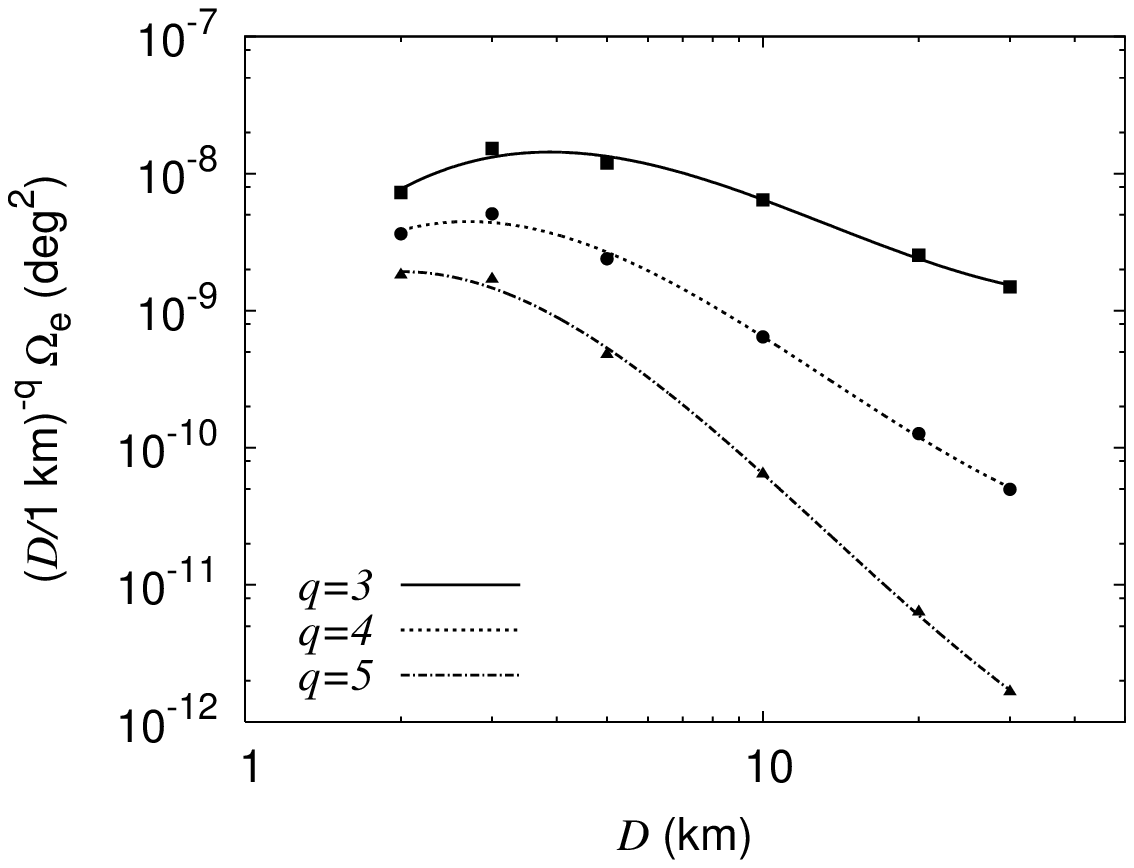} 
\caption{A plot of the the product of the power law size distribution
  $(D/\rm 1~km)^{-q}$ and $\Oe$ versus diameter $D$ at 100~AU, for
  several values of $q$. Note that $D^{-q}$ decreases with $D$, and
  $\Oep$ increases with $D$. The product of these two terms is
  an indicator of the sensitivity of the TAOS system.}
\label{fig:sensitivity}
\end{figure}

The sensitivity of the TAOS system to the size distribution is shown
in \fig{fig:sensitivity}. For $q = 3$ the the survey has the maximum
sensitivity for objects with $D = 3$~km at $\Delta=100$~AU, and the
diameter of maximum sensitivity decreases as $q$ is increased. Note
that the sensitivity drops off for larger objects, justifying our
choice of $\Dmax = 30$~km.

% Figure 11
\begin{figure}[b]
\plotone{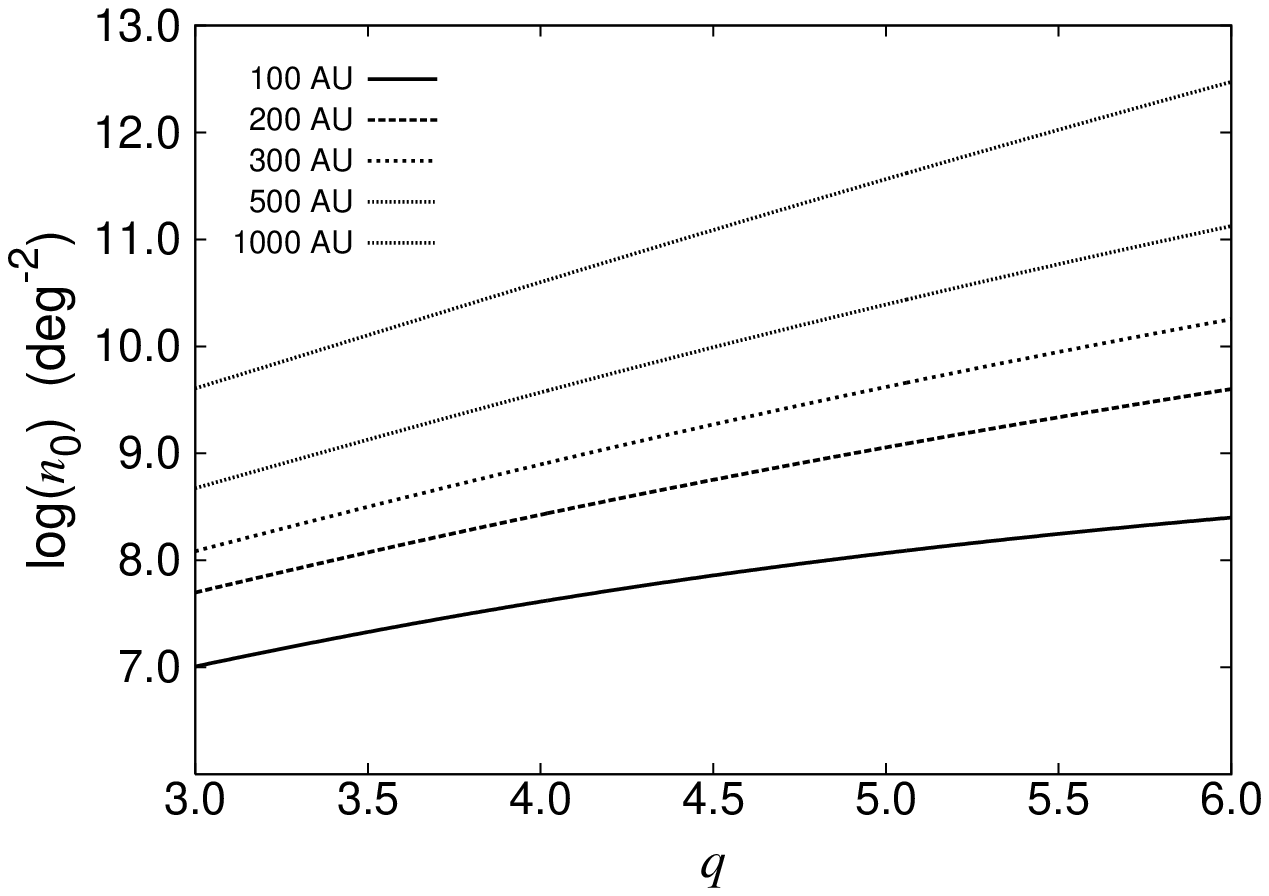} 
\caption{The upper limits of number density with objects larger than
  1~km for various q and $\Delta$. Any model with $n_0$ above the
  relevant line is ruled out at the 95\% c.l.}
\label{fig:nsedna}
\end{figure}

We are now ready to bring in observational results to set constraints
on the Sedna-like population, including the number density and size
distribution.  The null detection in the TAOS data means that any
model which predicts $\Nexpp > 3$ can be ruled out at the 95\%
confidence level (c.l.). Resulting upper limits on $n_{0}$ as a
function of $q$ are shown in \fig{fig:nsedna}.  Most of our target
fields have ecliptic latitudes $|b| < 10^\circ$, with more than 93\%
of the data collected in fields with $|b| < 3^\circ$, so the upper
limits we derived on $n_0$ are valid along the ecliptic, assuming that
the size distribution is independent of ecliptic longitude.

% Figure 12
\begin{figure}[b]
\plotone{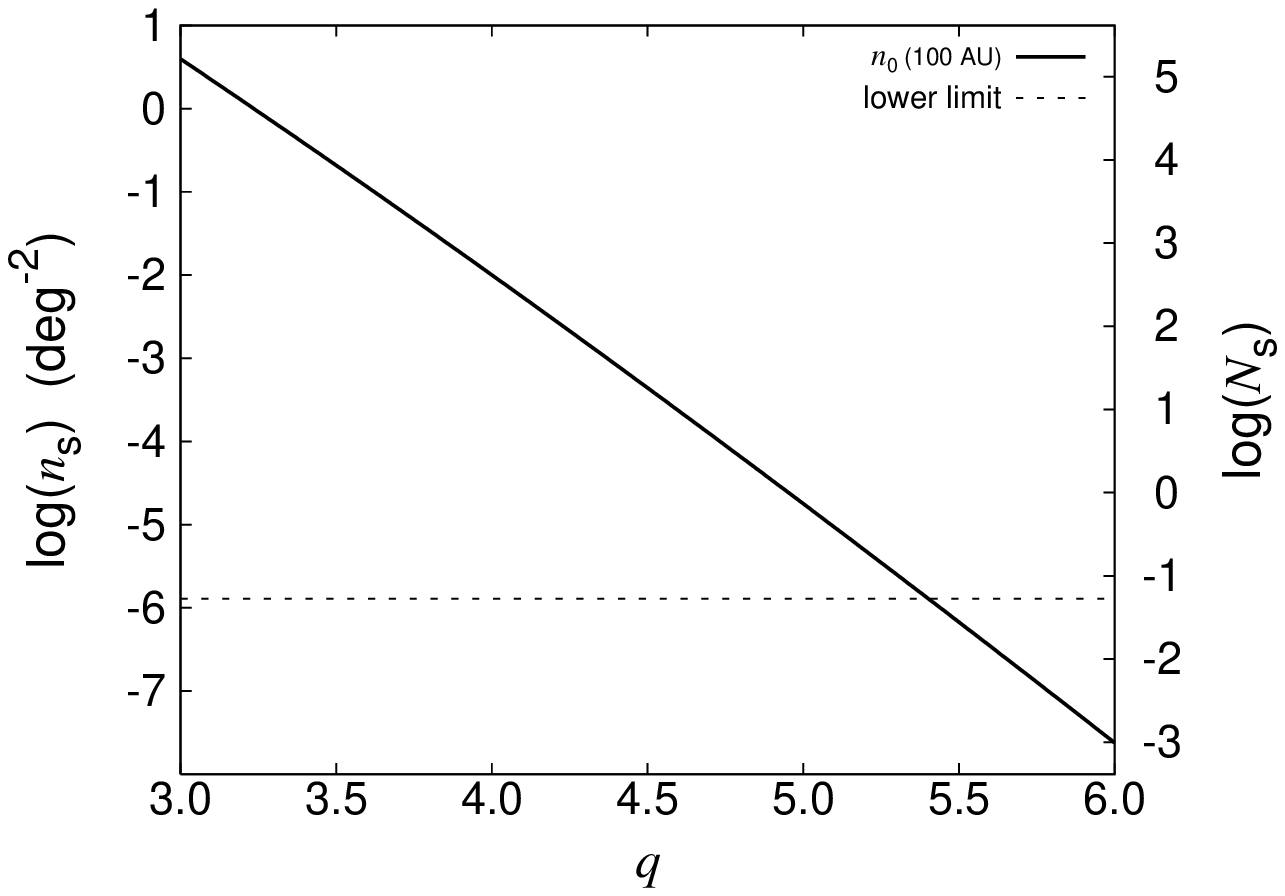}
\caption{Upper limit on $\ns$ as a function of $q$ at 100~AU. The
 diagonal line is the 95\% upper limit set by the TAOS survey at
 100~AU. The right axis shows the corresponding upper limit on the
 total expected number of objects larger than Sedna $(\Ns)$ in the
 whole sky,assuming an isotropic distribution. Given that one Sedna 
 actually exists, we can also set a \emph{lower limit} on the surface density.
 If our assumption of a power-law model is correct, we can exclude size 
 distributions where $\Ns < 0.05$ (dotted line) at the 95\%~c.l., 
 which corresponds to a value of $\ns<1.2\times10^{-6}$~deg$^{-2}$.} 
\label{fig:sedna}
\end{figure}

We know very little about the size distribution of objects in
Sedna-like orbits; Sedna is the only object that has been found.
Given the assumed power-law size distribution, we calculated the upper
limits on the density $\ns$ of objects larger than Sedna by
integrating \eqn{eq:dndD},
\begin{equation}
  \ns = n_0 \left(\frac{\Ds}{1~\mathrm{km}}\right)^{1-q},
\end{equation}
where $\Ds = 1600$~km is the diameter of Sedna.  This upper limit on
$\ns$ vs. $q$ at 100~AU is shown in \fig{fig:sedna}.  Because one
Sedna has been found near 100~AU, we can exclude any model that
predicts fewer than 0.05~objects with $D\geq\Ds$ at the 95\% c.l.
That is, with the power-law size distribution, any size distribution
which yields $\ns<1.21\times10^{-6}$ deg$^{-2}$ is excluded. An
interesting consequence is that slopes $q > 5.4$ are also excluded. A
larger slope means either too few large-sized objects (D=1600~km) to
be consistent with the existence of Sedna, or too many small-sized
objects ($>$ 1~km) to comply with the null detection by TAOS.

\begin{figure}[b]
\plotone{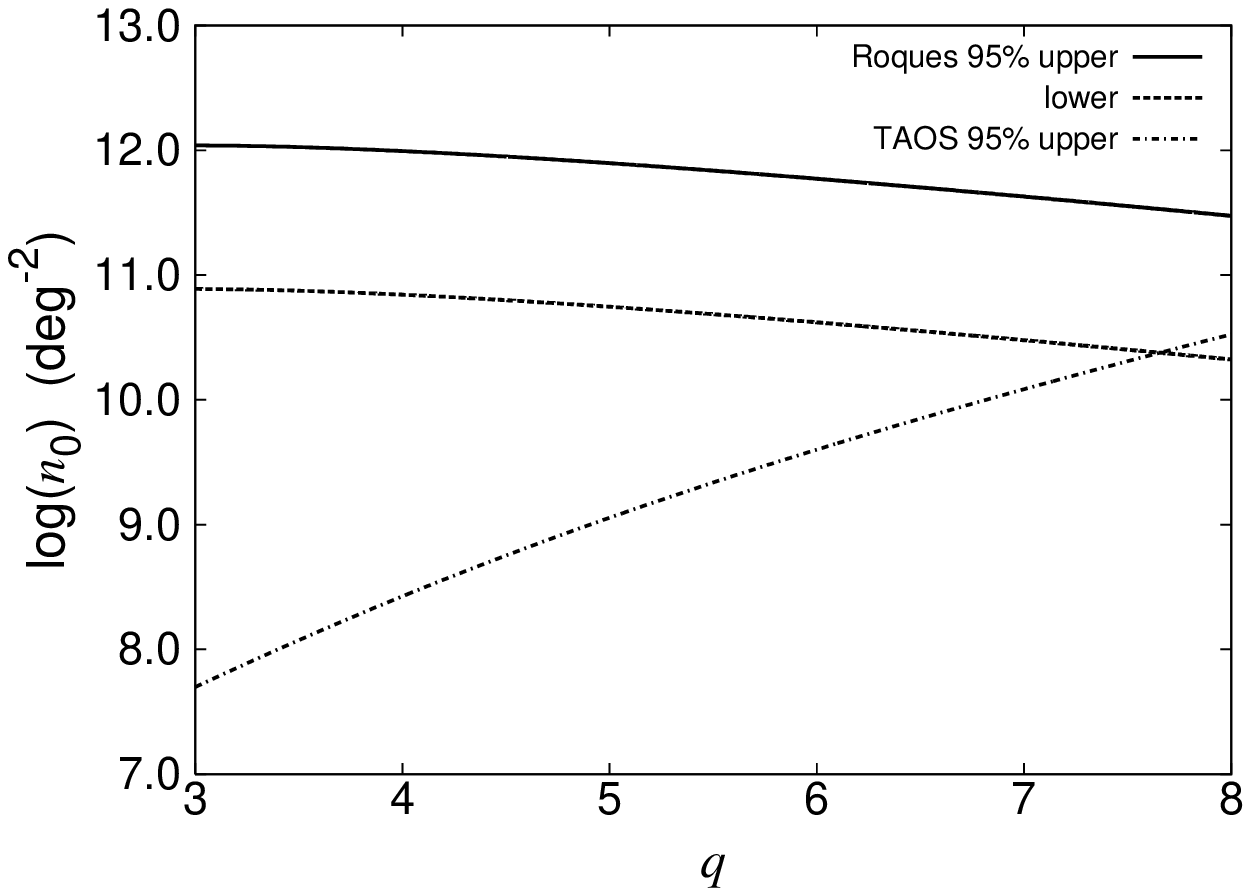}
\caption{The 95\% upper and lower limits on number density were
  calculated based on two $\sim600$~m objects found by
  \citet{2006AJ....132..819R} and $q$ from 3.0 to 8.0, along with the
  95\% upper limit set by TAOS survey at 200~AU.}
\label{fig:roques}
\end{figure}

\citet{2006AJ....132..819R} reported the possible detection of two
occultation events consistent with sub-km objects near 200~AU. The
resulting 95\%~c.l. upper and lower limits on $n_0$ vs. $q$ are
plotted in \fig{fig:roques}, along with the 95\% c.l. upper limit set
by TAOS survey at 200~AU.  Again, if the assumption of a power law
size distribution is valid, the interpretation of these two events as
true occultations is consistent with the TAOS upper limit at the
95\%~c.l. only if $q > 7.6$.  Note that this result is valid for any
population of objects (e.g. Sedna-like objects, the Scattered Disk or
Inner Oort Cloud) in this region.

% Figure 13

%%%%%%%%%%
\section{Conclusion}
We have analyzed the first two years (2005 February to 2006 December)
of TAOS data to search for objects in Sedna-like orbits. With the null
detection of objects in this region, we are able to set upper limits
upon the number density of objects at different distances. Given that
one Sedna has been found, we have also placed lower limits on the
number density of objects at 100~AU. We also show that the candidate
events reported in \citet{2006AJ....132..819R} are inconsistent with
the TAOS data at the 95\% c.l. if the size distribution at
$\Delta=200$ AU has slope $q < 7.6$. We plan to modify the detection
algorithm to facilitate an exact calculation of the statistical
significance of candidate events, and we will subsequently complete
the analysis of the second TAOS data set (January 2007 to December
2008) and report its results in a future paper. We will also expand
the analysis to include limits on the populations of Scattered Disk
and Inner Oort Cloud objects.

\acknowledgements Work at NCU was supported by the grant NSC
96-2112-M-008-024-MY3. Work at the CfA was supported in part by the
NSF under grant AST-0501681 and by NASA under grant NNG04G113G. Work
at ASIAA was supported in part by the thematic research program
AS-88-TP-A02. Work at Yonsei was supported by Korea Astronomy and
Space Science Institute. Work at LLNL was performed in part under
USDOE Contract W-7405-Eng-48 and Contract DE-AC52-07NA27344. Work at
SLAC was performed under USDOE contract DE-AC02-76SF00515.

\bibliographystyle{apj}

\end{document}